# Emerging AI Technologies Inspiring the Next Generation of E-textiles

FRANCES CLEARY[1,2], WITAWAS SRISA-AN[3], David C. Henshall.[1] and Sasitharan Balasubramaniam [3]

[1]Physiology and Medical Physics, RCSI Uni- versity of Medicine  Health Sciences, Dublin, Ireland
[2]Walton Institute, South East Technological University, Waterford, Ireland
[3]School of Computing, University of Nebraska, Lincoln, United States

**ABSTRACT** The smart textile and wearables sector is looking towards advancing technologies to meet both industry, consumer and new emerging innovative textile application demands, within a fast paced textile industry. In parallel inspiration based on the biological neural workings of the human brain is driving the next generation of artificial intelligence. Artificial intelligence inspired hardware (neuromorphic computing) and software modules mimicking the processing capabilities and properties of neural networks and the human nervous system are taking shape. The textile sector needs to actively look at such emerging and new technologies taking inspiration from their workings and processing methods in order to stimulate new and innovative embedded intelligence advancements in the etextile world. This emerging next generation of Artificial intelligence(AI) is rapidly gaining interest across varying industries (*textile, medical, automotive, aerospace, military*). It brings the promise of new innovative applications enabled by low size, weight and processing power technologies. Such properties meet the need for enhanced performing integrated circuits (IC's) and complex machine learning algorithms. How such properties can inspire and drive advancements within the etextiles sector needs to be considered. This paper will provide an insight into current nanotechnology and artificial intelligence advancements in the etextiles domain before focusing specifically on the future vision and direction around the potential application of neuromorphic computing and spiking neural network inspired AI technologies within the textile sector. We investigate the core architectural elements of artificial neural networks, neuromorphic computing (*2D and 3D structures*) and how such neuroscience inspired technologies could impact and inspire change and new research developments within the e-textile sector.

## I. INTRODUCTION

SMART clothing traditionally refers to a garment with the capability to enable/disable a function such as monitoring a person's physical condition [5], whereas an e-textile provides an added layer of intelligence such as connection to a peripheral or embedded electronic device into the garment or fabric, providing added value to the person that wears the item. Smart clothing leveraging embedded intelligent e-textiles with computational and memory capabilities are foreseen as the next big market mover in the Internet of things space. Demand for usable and wearable technology is constantly growing and the expected market traction was forecasted etextiles growth from *2981 Million Dollars (2022) to 8508.1 Million Dollars (2028) stated by Absolute reports*. Three generations of smart textiles have evolved over the years (1) *1st generation* of smart textiles: little integration between the electronics and the textile (2) *2nd generation* of smart textiles: evolved with the adaptation of traditional textile fabrication methods to include additional functionality. eg sewn in conductive thread into textiles. (3) *3rd generation* integration of electronic sensing properties into textile materials. What will the next generation of etextiles and smart clothing look like. Advancements of ICT technologies (artificial intelligence (AI)) intertwined with nanotechnology (*nano-textiles and wearable sensing nanomaterials*) are key enabling drivers of the next generation of smarter and more advanced etextiles driving AI inspired computing fabrics. Rapidly Advancing ICT technologies such as artificial intelligent spiking neural networks (SNN) and neuromorphic computing core technologies, computational capabilities as well as their architectural structure bring the potential to inspire new and fresh innovations in this domain.

The textile sector has been experiencing a digital transformation predominately within its textile manufacturing processes and production industry, where AI-enabled technologies are being adopted for production line fabric inspection and defect detection, enhancing output quality [3] [4]. [14] provides a survey of state of art technological interventions that meet automatic fabric defect detection aligning to the industry 4.0 initiative, detailing traditional(statistical methods, structural methods, model based methods) as well as learning based methods (machine learning or deep learning). Intelligent clustering and classification techniques adopted and utilised in the textiles industry are summarised in [13], highlighting both supervised and unsupervised learning types supporting production planning, fabric fault detection, performance and predictive models. In this paper we will examine in more detail artificial intelligence ICT advancements focusing specifically on neuromorphic computing and spiking neural network artificial intelligence, assessing their architectural structure, vision, capabilities and how these elements could be of relevance to inspire future research advancements in the e-textiles sector. Healthcare is emerging as one of the key sectors where e-textiles and new advances in artificial intelligence (AI) driving embedded textile intelligence and on-body computation, can be leveraged and utilised in the near future both within a clinical environment such as a hospital and also support enhanced remote monitoring of patients from the comfort of their own home. [37] details a smart garment *MyWear* that monitors and collects physiological data (*muscle activity, stress levels and heart rate variations*) processing the data in the cloud and providing predictions to the user based on abnormalities detected. Such e-textile applications and services utilising these AI technologies bring the added value of a more effective real time monitoring and analysis for varying health conditions such as cancer care, cardiovascular and neurological disorders, leading when required to early interventions as critical health concerns are detected. The '*Internet of Smart Clothing*' [12] pushes the boundaries around smart garment inter communication, their interaction with environmental objects and how they actively communicate with remote servers for the provision of advanced services. The next generation of smart clothing and e-textiles brings more intelligent embedded technological layers than before and hence has requirements for more flexible, modular, integrated, seamless and usable functionality to meet end user needs.

The rest of the paper is organized as follows: Section 2 provides an overview of advancements in nano technology and nano materials. Section 3 details AI intelligence currently impacting the textile sector and highlights four core technical properties required to be considered linked to data flow communication and process methodology. Section 4 discusses the core architectural properties of a spiking neural network and section 5 discusses the core architectural properties of neuromorphic computing with the objective to provoke thought around how inspiration can be taken from these elements and actively fed into next generation of etextiles.

## II. NANOTECH TEXTILE RESEARCH ADVANCEMENTS

Nanotech fabric that involves the integration of electronics with textiles are pushing the boundaries in an exciting new phase of interactive smart garments [39][38]. Nanotechnology involves the manipulation of individual atoms and molecules in order to create nano-scale machines that open the potential for a variety of applications across many sectors (*Sensors, Repel liquids, sensor display technology, odour control, fabric strength, wrinkle resistant*). [9] highlights the advantages of Wearable memories and Computing devices (WMCs) and recent advances in nanotechnology and materials science. The link between WMC and the human brain could enable fast operation along with interface complexity, directly mapping the continuous states available to biological systems.

Advances in micro-sensor technology and nano materials are feeding new and disruptive innovations around how we wear and actively engage with our clothing. Nanomaterials are enabling the next generation union of textiles, electronics and information processing [Fig 1]. Through this combination or hybrid approach this opens the door to integrate new capabilities such as antimicrobial properties or the integration of biological functions. Nanomaterials can be added during the fibre production phase or also during the finishing phase onto the fibre surface. The aim is to utilise nanomaterials in order to achieve enhanced flexibility, usability while also being more fashionable. Nanomaterials consist of various materials such as graphene, carbon nanotubes, ploymers, dielectric elastomers and composites. Based on various stimuli to such nanomaterials this creates different characteristic behaviours that can in turn be utilised for varying applications specifically in the healthcare sector related to patient care applications. Biosensors that look at utilising nanomaterials integrates varying disciplines ranging from molecular engineering, material science, biotechnology/chemistry offering high sensitivity towards the recognition of various disease biomarkers at an early stage[40].

In 2004 researchers succeeded in isolating graphene sheets (exfoliated graphene). This two dimensional material consists of carbon atoms arranged in a hexagonal lattice. It is the thinnest material known in the world (*one atom thick*) and opens the door to multiple opportunities in the design and development of intelligent and smart garments and etextiles. Graphene is light, has greater elasticity and conductivity and brings the potential to replace synthetic fibres( polyester, nylon). Chemical sensing properties through graphene fibers can be introduced into textile based materials. *Cutecircuit* (*https://cutecircuit.com*) created a black dress that contained graphene. The dress had the capability to change colour in synch with the wearers breathings. This was completed using mini LED lights with graphene used to power the lights and as a sensor to record the wearers breathing. Graphene based composites have also been used in 3D printing technologies to investigate the development of highly stretchable and sensitive strain sensors. Such an application can be leveraged for breathing pattern monitoring and measurement [41].

Nanoparticles are emerging as diagnostic sensors. Textile advances at nanometer level encourage a new era of improved textile functions and capabilities for example enhanced communications, antibacterial, sensing capabilities of textiles or moisture absorbing. Garment communication through easily integrated textile antennas and interconnected integrated wearable textile systems are completed through the use of electroconductive fibres/yarns and embroidery techniques. Research in this area has been actively investigating material properties (flexibility, conductiveness, weight) leading to the emergence of nanoengineered functional textiles and the addition of nanoparticles to fabrics to increase textile properties and capabilities. (Ghadremani et al 2016) details the addition of silver nano particle in order to improve the static properties of a textile. (Liu et al 2010) highlights the addition of Ag (*Silver*) nanoparticles on sock garments that in turn limit the growth of fungi and bacteria, reducing odour and itching of the feet. Nanomaterial skin-like wearable sensors, flexible electronic substrates are the focus of development at industry level. There has been ongoing attempts to mimic human skin flexibility and stretch-ability in the etextile world. Sample Smart skin developments include

- (Tianzhao Bu et al) developed a stretchable triboelectric-photonic skin with multidimensional tactile and gesture sensing.
- (Liu et al) completed a review on Lab-on-Skin, in particular flexible and stretchable electronics for wearable health monitoring.
- (Yang Et al) presents a novel structured fibre sensor embedded in silicone for the precise measurement of low-pressure in smart skin.

Such nanotechnology, nanomaterial and smart skin advancements are attracting alot of attention both within the research world and also with businesses as they see the economical impact and the potential end user commercial applications of promise that are emerging. In parallel the next generation of artificial intelligence is seeing the emergence of Spiking Neural Networks(SNN) and neuromorphic computing, that will no doubt in the future trigger research directions focused on a fusion of nanotechnology AI-driven embedded E-textiles innovations. The remainder of this paper will delve further into the AI (*SNN*) and neuromorphic engineering technologies mapping and highlighting key architectural aspects to be considered further within the textile and smart garment research and innovation domain.

## III. ARTIFICIAL INTELLIGENCE IMPACTING TEXTILE SECTOR

Etextile research domain experts, suppliers and manufacturers are starting to investigate at a deeper level the impact AI and machine learning technologies can have across varying sectors [6]. Digital transformation through the use of artificial intelligence is currently impacting the textile industry through the creation of a more sustainable digital supply chain and smart intelligent textile manufacturing optimization(*production planning and operational process management*) right through to fabric defect identification [2], pattern inspection analytics and much more.(Chaudhari et al) provide an insight into the application of Artificial neural network (ANN) in fabric engineering, focusing on the application of ANN in the textile domain to support fabric classification, fabric wrinkle, automatic recognition of fabric patterns and fabric comfort. How such enhanced and smart technologies can aide the textile industry towards a sustainable and circular economy is of high priority and gaining a substantial amount of attention. Textile fabric based design software is also seeing the adoption and usage of AI based software tools in pattern design, making and cutting, providing a superior level of tools with inbuilt 3D visualisations features [1].

The adoption and use of AI within fabric based textiles requires a structured and methodological process taking into account technical properties of importance along with the end user functional and form factors (Giuseppe et al)[13]. Such key etextile embedded functionality technical properties that are required to be considered when contemplating taking inspiration from advanced AI technologies into a fabric based environment include

- *Fabric based textile **computational considerations** (Data Acquisition and Data Processing)*
- ***Monitoring and measuring** capabilities and techniques suitable for a fabric based environment.*
- *Suitable **communications methods** and techniques.*
- ***Energy considerations** of relevance for textile-driven intelligence .*

All these elements directly relate to the data flow communication and processing methodology. In this paper we will adopt these elements in order to map across new AI technology properties to the textile domain within section 4 and 5. Currently within the healthcare medical sector such technical functional properties are applied and demonstrated through intelligent e-textiles for patient centric garment-based wearables[19]. Such considerations in the design and development enable the possibility to gather human monitored health related datasets such as Electromyography (EMG) or Electrocardiography (EKG) data-sets through textile based sensors in wearable garments. Such data collection type garments and textiles need to be adaptable to the users needs for ease of use at varying levels. Through the active collection of these datasets, this then allows for the transmission, processing and extraction of key analysis and results for effective decision making, this is aided through the development and implementation of intelligent algorithms. The use of Artificial intelligent based algorithms and techniques enables an era of intelligent textiles utilising real time and accurate data knowledge in dynamically changing healthcare monitoring type environments.When considering such embedded AI in a textile environment, there is a need to also look at the state of art of current AI formal methods and how they must be advanced to adapt. Key challenges and questions arise that will require further investigation and research as technology advances and emerges in this space.

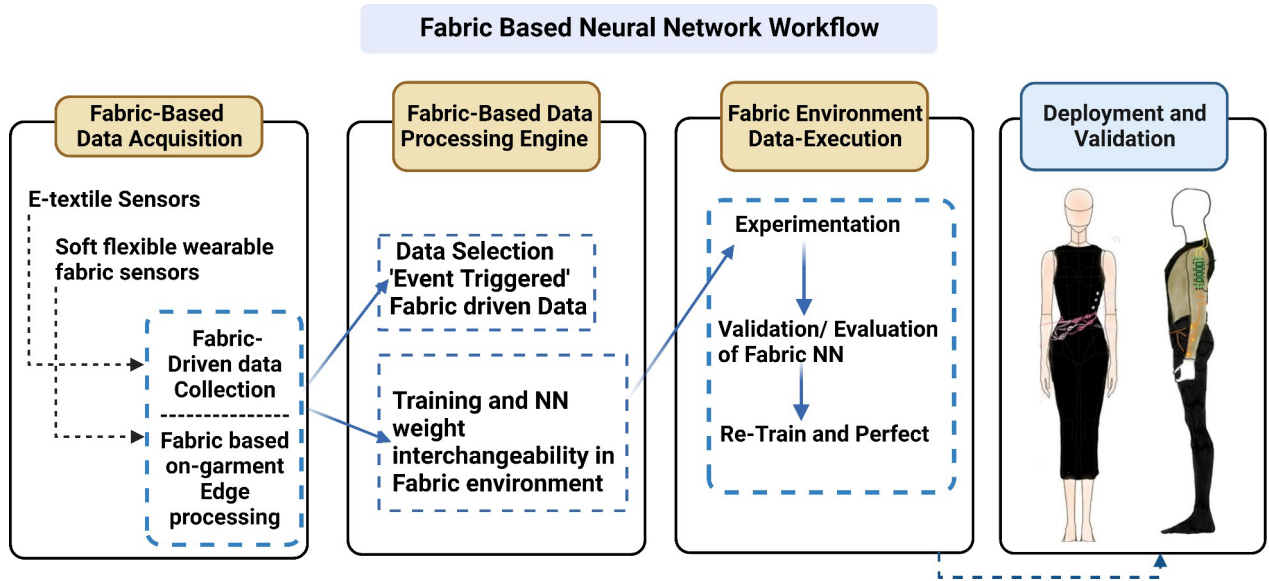

FIGURE 2. Fabric based Neural Network Workflow.

Pushing the boundaries around the use of conductive threads and embedded smart wearable intelligence and functionality [48] were successful at their attempts to store data in fabric, this was completed through the magnetic properties of conductive thread and this allowed for applications such as utilising a garment to store a passcode to open a door to gain access to a building through the swipe of the garment arm that holds the passcode.

We must ask ourselves what future fabric-based AI algorithms will look like, how will they be developed and integrated in a functioning manner in the very core of a fabric and fibre environment. Not only will new Fabric AI design fabrications emerge but also new fully fabric driven data acquisition and data processing driven approaches potentially adopting or inspired by new AI best practice, standards and methodologies. Key features of fabric-based AI algorithms need to consider during implementation speed (response time), processing capabilities, complexity/size and also learning. [Fig 2] provides an overview of the key components for consideration for a textile driven AI workflow facilitating a fabric based data acquisition, data processing fabric-based engine, execution in a fabric environment and also deployment.

There is an emerging research interest around the application of artificial intelligent technologies in smart wearable garments and the internet of smart clothing. Research questions are being investigated around how such AI intelligence can be embedded into the very core of a fabric environment, with AI intelligence applications seamlessly embedded invisible to the human eye. [6] provide a comprehensive review of advancements in Smart textile integrated microelectronic systems,highlighting the core properties of importance 1) flexibility allowing for effective drape on 3D curvilinear surface such as a human body and 2) structural transformation of textiles resulting in low fibre strain and fabric life cycle longevity. [11] convey a vision of moving from fibre devices to *fabric computers*, where the fabric fibers have inbuilt capability to perform sensory,storage,processing and powering capabilities providing a fabric based computing environment. Such a powerful fabric fibre based processing capability enables the execution of fabric based programs that can activate fiber sensors, processing and storing data within the fabric computer. Work has been ongoing around the development of such new fibres with specific focus on scalable processes using thermal drawing, melt spinning, coating to provide fibre structures that can house and deliver computing functionality[7][8][10]. Investigative methods into the digital fabrication of fibres is being researched where inbuilt func-tionality provides in-fibre storage programs, data storage, sensors and digital communication [11], such a fabrication structured process proposes uniform placement of discrete in-fibre electronic devices that will carry out such functionality. Researchers are actively thinking outside the box about new and potentially disruptive innovative ways to fuse AI with etextiles moving away from the traditional textile world as we know it. This is sparking a renewed interest in this domain and shows promise of real impact across multiple sectors.

## IV. SPIKING NEURAL NETWORK PROPERTIES

Artificial neural networks are seeing the emergence of new spiking neural network (SNN) technologies that simulate functionality using electronics components replicating and mimicking human brain biological workings of neurons, synapses and neural networks. Core architectural properties of an SNN include

- Neurons that emit a spike once a set threshold has been met.
- Learning in the neural network is completed by altering the synaptic weight. Random weight change algorithm is one of the most adopted and simple algorithms used during the learning phase. For this algorithm the correct output is known and the error increased or decreased as required.
- Results obtained depends on the neuron spiking activity and also the neural node inputs.

[15] detail and explain a single neuron level, giving 1D neuron model examples such as Leaky integrate and fire (LIF) [16],the Spike Response Model (SRM)[17] as well as more complex and biologically feasible artificial neurons such as the the Hodgkin and Huxley model[18]. We will now assess the core computational, monitoring /measurements, communications and energy of such SNN, highlighting structural and processing paradigms inspired by the human brain and the potential they could bring to the etextiles domain. [Fig 2] details the adoption of this data flow communication and processing technology methodology to the etextile domain highlighting a fabric based workflow of relevance towards the implementation, validation and deployment of fabric driven AI ( neural network ) intelligent etextile wearables.

### A. COMPUTATIONAL ELEMENTS OF SNN ENABLING FABRIC-DRIVEN DATA ACQUISITION AND PROCESSING

Data acquisition refers to the methodology and process of acquiring data and performing analysis in order to interpret it. This involves the use of varying techniques and tools used to sample data, convert the data into a format that can in turn be used for further analysis and processing. From a neural network point of view, we will now investigate further the main computational elements with a focus on highlighting architectural aspects of importance for consideration in a textile environment 1) *Artificial fabric neurons* 2) *Artificial fabric synapses* 3) *Artificial fabric neural networks* required to perform data acquisition and processing and the potential for adoption into a fabric environment.

#### 1) Etextile Artificial Fabric Neuron

Neurons are the core building blocks of neural networks. The workings of a neuron include synapses represented by weights, a threshold and an output spike that in turn resets the neuron. Each neuron has a membrane potential. This membrane potential is the equivalent of a voltage and when that voltage passes a defined threshold a spike (*or action potential*) is emitted and hence this generated spike is the method by which one neuron communicates to another neuron in a SNN. Taking these aspects, how can we begin to consider a fabric based neuron, its workings to replicate not only an individual neurons functionality but having the capability to be extended to implement multiple interconnected neural nodes in a fabric environment. To work towards such a goal, we have to delve into the artificial electronic neuron representations currently defined and that could inspire future fabric based neural implementations. Here we will consider the Leaky integrate and Fire Model and the Hodgkin and Huxley Model. Further enhancements to these models and other models exist, but this is outside the scope of this paper. *Leaky Integrate and Fire Model (LIF)*: The most simplistic model of a neuron is the *Leaky Integrate and Fire neuron* artificial electronic circuit based on the logic that if the spike (driving current) goes beyond the defined threshold then the neuron emits its own spike and resets. The model operates based on a resistor and capacitor (RC circuit) [Fig 3]. Limitations of the LIF model is that no memory of spiking activity is retained as the membrane potential is reset after each spike. We need to consider how such a LIF model could be replicated to produce an event driven spiking neuron in a fabric environment leveraging the leaky integrate and fire neuron model. Soft fabric based resistors can be fabricated using conductive thread (Zigzag machine stitch to create) and ongoing research is active around the development of textile based capacitors[20].[51] investigate methods for the capac-ity increasing of textile capacitors for planar and sandwich type textile capacitors using hybrid conductive threads and conductive textiles. Experimental research has also been on-going into the development of wearable fabric Brain enabling on-garment edge-based sensor data processing inspired by SNN architectural techniques and LIF model [56]. Such advancements will open up options and new methods to work towards functional fabric based neural nodes based on an RC circuit and LIF model, capable of processing 'event' spike driven activity replicating a neural node, that can be extended into a basic working event driven spiking neural network.

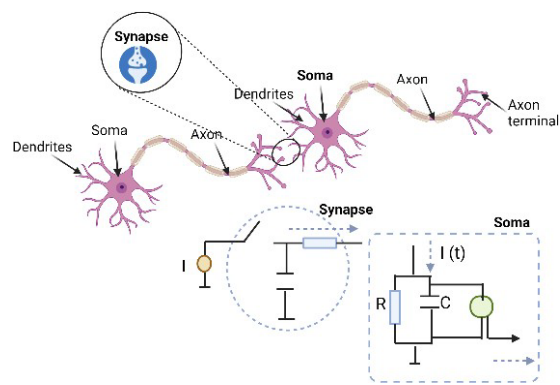

**FIGURE 3.** Simple Integrate and Fire artificial (RC) neural node

Additional embedded transistor gated circuitry is required in order to implement the threshold level $V_{th}$ for the membrane potential, when this threshold has been reached this produces an action potential spike. [22] detail an organic field-effect transistor that is realized on a flexible film that can be applied, after the assembly, on textiles. [23] showcase a fully inkjet-printed 2D-material active heterostructures with graphene and hexagonal-boron nitride (h-BN) inks, which are used to fabricate an inkjet-printed flexible and washable field-effect transistors on textile. Such advancements in textile based FET's bring the capability to incorporate textile threshold level gates enabling the possibility of varying threshold levels for the various neural nodes in a fabric spiking neural network.

*Hodgkin and Huxley* neuron model is a more complex model to replicate the generation of an action potential of a neural node. The model can describe the time behavior of the membrane potential and currents through potassium (K) and sodium (Na) channels using differential equations[Fig 4. They were able to observe the generation of action potential as well as the refractory period [21]. In this circuit the capacitor is representative of the cell membrane, the circuit has variable resistors that represent the voltage-dependent K+ and Na+ conductance's and there is also a fixed resistor representing the voltage-independent leakage conductance. This model has three power batteries for the reverse potentials for the corresponding conductance's. Generation of this model in a fabric environment would require the textile resistor and capacitor as well as the requirement for a variable resistor interconnected using conductive thread. To date variable textile resistors have been created in the form of fabric based potentiometers.Such fabric potentiometers contain a conductive wiper function as well as a resistive track where its ends has measurement points included. The conductive wiper acts as a means to set and measure a variable resistance through adjustment of the sliding wiper. [49] detail and demonstrate a zipper based potentiometer. Other variable resistance elements that could be considered to produce such a variable resistor include include Eeonyx Stretchy Variable Resistance Sensor Fabric (*Adafruit*) that can be utilised to make soft sensors that are required to be movable and adaptable. Such stretch fabric sensors using stretchable conductive fabric enables changes in resistance when stress is applied.Further research is required into how textile and fabric based variable resistance can be leveraged and potentially be utilised in the design of a fabric neural node.

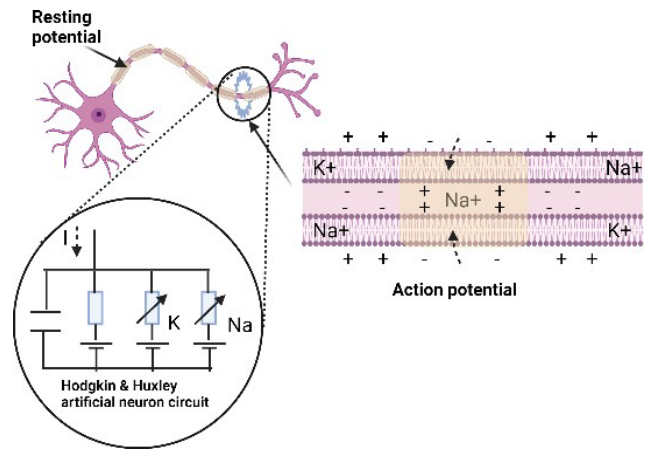

**FIGURE 4.** Hodgkin and Huxley electronic Artificial neural node

Key textile components are in existence to enable the creation of a fabric base Hodgkin Neural Node, how such elements can come together from a design perspective to produce a working neural node is the key challenge here in order to produce a working Hodgkin and Huxley Neural node.

2) Etextile Artificial Fabric synapse

Current learning in SNN are dependent on the capability to alter and interchange the synapse weights for each of the neural network nodes. When a neuron threshold is reached it fires and produces an action potential. This is a result of the sum of excitatory and inhibitory potentials and these are connected to the neuron through the synapse. We refer to the synapse as a synaptic weight. This is the strength of a connection between two nodes in a neural network. [25] focus on non-static neural networks at a signficant distance from each other and how through their implementation of an all-optical synapse stemming from wavelength division multiplexed visible light communications can overcome real-time weight adjustments. [52] investigated yarn coated with reduced graphene oxide (RGO) to produce two-terminal memristor-based artificial synapses suitable for use in wearable neuromorphic computing systems. [53] researched the design and development of a one-dimensional organic artificial multi-synapses enabling electronic textile neural network for wearable neuromorphic applications, where the multi-synapses comprising of ferroelectric organic transistors fabricated on a silver ($Ag$) wire.

To replicate a SNN in a fabric environment we need to consider the functionality required for the synapse and how to embed this in a workable manner into a fabric environment. As the weight influences the firing of a neuron, in a SNN, at a basic level this can be replicated by embedding the option to be able to connect and interchange from one conductive thread-based resistor in a fabric environment to another conductive thread resistor. Such textile resistors can in turn be utilized as synaptic fabric-based weights. Further advancements with the introduction of memristors as synaptic weights are emerging and need to considered from a fabric synapse implementation point of view, these will be covered in section 5 neuromorphic computing properties.

**TABLE 1.** Types of Neural Network more relevant to SNN's

| SNN Type | Data Flow | Memory | Prediction | Weights |
|---|---|---|---|---|
| FeedForward | Input to Output | No Memory | Poor | Weight matrix inputs, produces output |
| Recurrent | Looping Information cycles | Has Memory-learns | Good | Weights applied current and previous inputs |

3) Etextile Artificial Fabric Spiking Neural Network

There is multiple types of neural networks perceptron, multilayered perceptron, feed-forward, recurrent, fully connected, Convolution, Radial Basis Functional,Long Short-Term Memory (LSTM), Sequence to Sequence Models and Modular Neural Network. Table 1 provides an overview of a Feed-forward and re-currant neural network properties of relevance when considering a SNN type to adopt and conform to [42][43][44]. When designing an embedded neural network in a fabric environment key fabric and end user properties need to be accounted for including aesthetics, durability, comfort and maintenance. Based on the overall size of the SNN and the number of hidden layers it incorporates, this deciphers the number of textile artificial neural nodes,synapse interconnections and interconnected fabric multi-layers required.

We can then begin to investigate the best possible design, layout and functionality integration methods around how to accommodate and embed into a fabric environment.

B. SPIKING NEURAL NETWORK MONITORING/ MEASURING

A SNN can learn by supervised learning, where you have an input and an output variable and the algorithmic computation within the neural network learns from a training dataset, once an acceptable level of performance is achieved the learning stops. An unsupervised method in comparison has input variables but no output variables that are used to support training and learning of the neural network. [24] present an overview of the varying training methods for SNN's such as conventional deep networks, constrained training , spiking variants of back-propagation and variants of Spike time dependant plasticity (STDP) in order to categorise SNN training methods and also highlight their advantages and disadvantages. Creating such a training or learning process in a fabric etextile is a challenge that has not yet been achieved.

The use of nanotech is the obvious initial best approach to attempt to embed such a learning element into an intelligent fabric garment.

Depending on the structure of a SNN, this identifies its classification. For this paper we will focus on a fully connected multi-layer neural network. Such a multilayered Spiking neural network consists of multiple layers of artificial neural nodes (usually has three of more layers and utilizes a nonlinear activation function). From a design perspective in a fabric environment once we have identified the core textile components required to implement a working neural node as well as a functional method for interchangeable synaptic weights, the next step is to progress towards the identification of a most practical and feasible fabric-driven design in order to incorporate multi-layers, their interconnections and how to be capable of validating and modifying during the execution phase.

Several layers of fabric woven and stacked produces a multilayer fabric, secured together with connecting yarns in a third (Z direction) dimension. Such woven fabrication techniques along with layered and interwoven fabric manipulations are design options that need to be assessed to identify suitable and best practice design for the development of SNN technical textiles [Fig 5]. Weaving multiple layers in a fabric provide the opportunity to embed neural network nodes and interconnected neural networks in an embedded fabric environment . Research into best approach ,best methodology to adopt and also core components and their re-usability still remain under investigation, but as textile components and intelligence along with nanotechnology advances, new opportunities are emerging pushing towards this vision of a Fabric AI Driven intelligence. In section 5 we will delve a little further into 2D/3D stacked layered techniques taking inspiration from neuromorphic computing and advancements here.

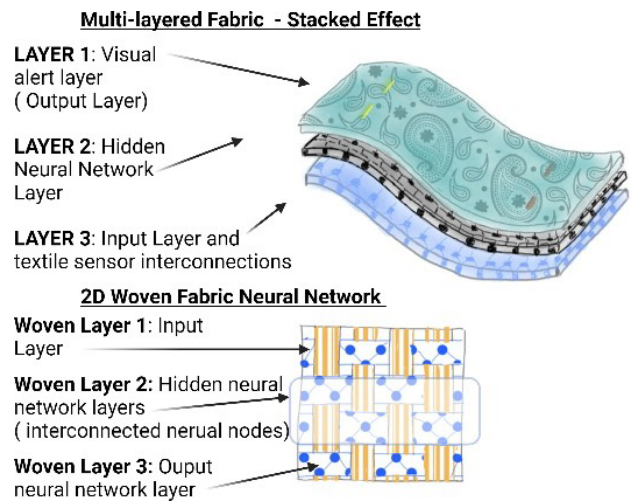

**FIGURE 5.** Multilayered conceptual design approaches inspired by stacked neural networks

**TABLE 2.** Neuronal Information Encoding Types

| Coding Type | Details |
| --- | --- |
| RateCoding | Rate of spikes in a set time interval. Can be used at single neuron level or interpretation spike trains. |
| BinaryCoding | A neuron is active or inactive in a set time interval. It fires 1 or more spikes within that timeframe. |
| Timetofirstspikecoding | Method is used to encode information for fast responses (in milliseconds). it is based on first-spike patterns). |
| FullyTemporalcoding | Precise timing of all spikes generated are part of the encoding. Timings are important. |
| PhaseCoding | Convert inputs into binary representations (eg a '1' equals a spike generated). Information is added by assigning different weights to each bit represented in the phase |
| BurstCoding | A burst of Spikes (Short and High frequency of Train of spikes). Assessment of the number of pattern of spikes in a burst coding |
| Latencycoding | Number of spikes is not the priority, instead the time between an event and when the first spike triggers is important. Stimulus is encoded based on a rank-order where neurons in a group generate their first spikes |

### C. SPIKING NEURAL NETWORK COMMUNICATIONS

SNN drives the adoption of brain-inspired computing, providing not only fast but also a large substantial amount of event-driven data processing. An SNN neural computation and communication is defined through the generation of spikes enabling neurons to communication from one to another via such triggered spikes. Research is ongoing around the types of neuronal information encoding's.[29] summarise the signal encoding schemes for a spiking neural network. Neuronal encoding and decoding is the information and communication process where for example an external variable or stimulus triggers neural activity within the brain. Such stimuli (e.g touch stimuli) produce varying neural activity patterns in the brain.[27] provides an overview of neural coding with bursts and new methods for their analysis. [28] introduce spiking autoencoders with temporal coding and pulses, trained using backpropagation to store and reconstruct images with high fidelity from compact representations.

Classification of the spike train pattern and what this means, enables active decoding of the pattern. One such example is the classification of spike train through active matching of the spike train patterns to templates. Such a template would be a set word, meaning or result. Within an etextile world neural information encoding, decoding and the creation and validation of potential *Fabric SNN classification templates* mapping to external sensing embedded textile sensors linked to fabric based neural networks, could enable the communication and processing of fabric based SNN encoding. Such Fabric SNN classification templates could correspond to an alert notification raising awareness around a critical health monitoring scenario where such a template could be used in conjunction with a SNN fabric Smart garment to assess the health status and provide feedback to the wearer based on the use of such classification templates to raise alerts to the wearer as required. Core to the fundamental working of a SNN is the manner in which the network nodes interconnect and how information flows and communication between the nodes is enabled. Multiple artificial neural network types exist. [26] provides a comparative overview of neural coding in a spiking neural network with in-depth detail on rate coding, time-to-first spike (TTFS) coding, phase coding, and burst coding. Table 2 provides a substantial list of the types of neuronal information encoding techniques utilised to-date for consideration. It highlights their key elements for consideration when investigating the potential for Fabric-driven SNN neuronal encoding and decoding.

### D. SPIKING NEURAL NETWORKS ENERGY CONSIDERATIONS

Spiking neural networks bring the promise of enhanced energy efficiency. As an SNN is a dynamic system this suits more dynamic driven processes and applications. Research is ongoing to investigate how to effectively lower synaptic operations and hence the computational performance of the neural network. [50] focus on the optimization of energy consumption of SNN for neuromorphic applications through a hybrid training strategy that also accounts for energy cost stemming from the networks computations. From a structure and architectural point of view SNN have typically fewer neural nodes than more traditional artificial neural networks along with the fact that SNN can implement '*node connection pruning*'[Fig 6] in order to reduce processing power and improve overall the working functionality and energy efficiency of the SNN. [30] develop a pruning method for SNNs by exploiting the output firing characteristics of neurons, which can be applied during network training. [31] detail the process of pruning STDP-based connections as well as quantizing the weights of critical synapses at regular intervals during the training process. They validate a classification accuracy of *90.1* percent and show an improvement in energy efficiency. When implementing a SNN in a textile environment, the capability to be able to disconnect and reconnect fabric neural nodes needs to be considered in order to be able to *prune* the Fabric SNN in an efficient manner enhancing the fabric SNN energy operational functionality. From a computational point of view the SNN has the capability to operate more quickly due to the neurons sending spike impulses. As SNN's adopt temporal information retrieval this increases the overall processing time and productivity and hence has a very positive end impact on energy consumption in the SNN.

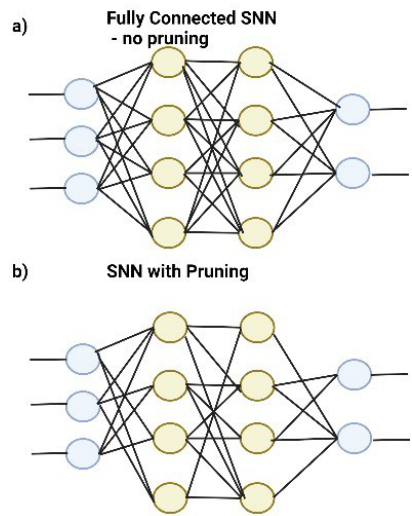

**FIGURE 6.** Difference conveying no pruning in SNN versus pruning in SNN

## V. NEUROMORPHIC COMPUTING PROPERTIES

Neuromorphic computing concept originated in the 1980's. Taking inspiration from computer science, mathematics to bio-inspired models of neural network. This emerging interdisciplinary research field has the potential to disrupt traditional computing methods and architectural implementations leading to a more centralized and combined memory and computational driven approach, moving away from the von Neumann architectural approach with separate memory and computing capabilities and high compute power needs. Such inspiration coming from the working of the human brain paves the way for new and more fault tolerant layered and parallel architectural designs and layouts.

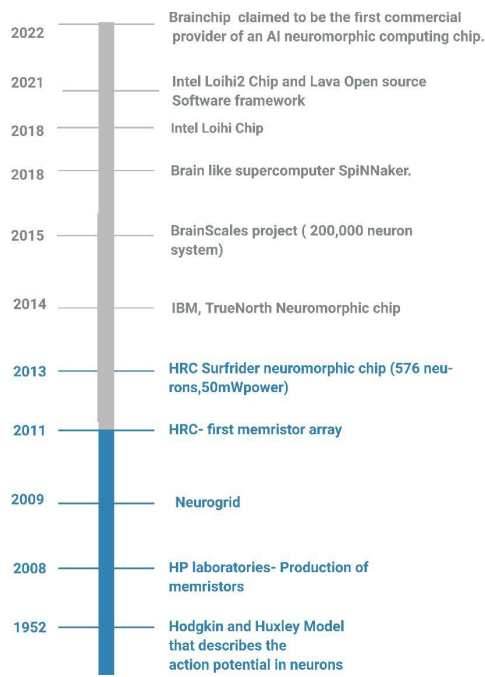

**FIGURE 7.** Timeline of technological advancements in Neuromorphic Computing

In order to understand what aspects of neuromorphic computing can inspire innovative advancements in the next generation of smart etextiles and on-garment edge based intelligence, we will first highlight the core key architectural elements of importance within neuromorphic computing and assess their potential within an etextiles domain. Neuromorphic computing core architecture is based on the concept of communicating through event driven spikes generated through simple processing structures represented by synapses and neurons. Ongoing research is pushing the production possibilities using complementary metal oxide semiconductor (CMOS) technology to develop neuromorphic spiking neural network hardware implementations [45][46][47]. Key properties such as *size, weight, low power consumption, and modular design (scalability)* are dominating the research areas of focus linked to such technologies. Over the years advancements in CMOS technology has driven smaller and more power efficient systems with the capability to mass produce. Such technology combined with advanced machine learning techniques has directly lead to the simulation and implementation of silicon based neurons, otherwise defined as neuromorphic computing.

[Fig 7] highlights advancements, with *Brainchip* (https://brainchip.com/) announcement in 2022 claiming to be the worlds first commercial producer of a Neuromorphic AI processor '*Akida*' that has the capability to mimic the working of the human brain and process data with high precision and energy efficiency. Akida being an event-based AI neural processor featuring *1.2 million neurons and 10 billion synapse*.

### A. NEUROMORPHIC COMPUTING COMPUTATION
1) Crossbar Array Architectural Properties

As neuromorphic computing moves away from the traditional Von Neumenn architecture towards a more focused in-memory computational architecture, the processing occurs inside the memory functionality elements, hence reducing data transfer time and energy. Hardware architectural design considerations need to take into account 1) *synapse interconnections between neural nodes* and 2) *how this can be implemented in order to complete a fully connected neural network*. From a hardware design perspective the cross bar array architecture has been adopted in neuromorphic computing in order to implement a full complement of interconnections required for to meet the neural network structure requirements. The cross bar array architecture includes a number of rows (word lines) and columns (bit lines) with memory devices interconnected between both the row and column. Advancements have been made through the development of resistive memory devices known as memristors (*one transistor and 1 resistor combination*). The operational functionality of the crossbar array is based in input current (*voltage pulse*) to selected rows which in turn activates selected columns via a voltage pulse, depending on the activation of varying cells in the crossbar array. For active cells in a particular row/column vertical line in the crossbar array, the sum of currents equals the output current, calculated using Ohms law and Kirchhoff's law. Research into memristor crossbar arrays for brain inspired computing neural networking has been investigated and summarised by [35]. [54] report a *64x64* passive crossbar circuit that demonstrates approximately *99* percent nonvolatile metal-oxide memristors, enabling the active storing of large neural network models on neuromorphic chips.

**TABLE 3.** Taking Inspiraton from SNN and Neuromorphic advancements, highlighting key considerations to feed into future etextile neural network research and prototypes

| Human Brain- NN properties/Neuromorphic Computing Properties | Inspired Etextiles Considerations |
|---|---|
| **Synapse**/Memristor: component that regulates electric current flow remembering the previous current flowthrough. | How to represent weights in a fabric environment to simulate synapse interconnections between fabric-based neurons. Conductive thread-based resistors, surface mount devices or nanodevices in a fabric environment to simulate the workings of the synapse weights. Interchangeability of these weights in a fabric environment need to be considered and generation of a new method and fabric-based process around how to design , develop and validate. How to embed in a workable manner a wearable memory aspect in a fabric environment, in a seamless functioning manner. What would a fabric based memristor look like, how could this be completed and validated in a fabric environment. |
| **Soma**/Neuristor: Device to capture the properties of a neuron, Spike or impulse generation when threshold reached. | Replication of a Neuron in a fabric environment.Take inspiration from current electronic artificial neurons (LIF, Hodgki Huxley).Considerations around how to create textile and fabric-based components to replace hard component elements (conductive thread-based resistors, textile-based capacitors). The aim being to investigate the most suitable way to create a working fabric neuron but with limited hard components instead using textile versions. |
| **Axon**/Circuit interconnections and signal conditioning. | In order to maintain the focus on a textile-based implementation the use of conductive thread (embroidered conductive thread neuron interconnections design pattern) as a means to easily reproduce and create such fabric neuron interconnections. |
| **Dendrite**/3D architectural design implementation, pattern detection and sub threshold filtering. | Taking inspiration from the 3D architectural design used in Neuromorphic computing , can this motivate and inspire new 3D fabric manipulation and 3D fabric layering type designs in a garment structure , to embed and interconnect such fabric driven neural network functionality. |
| **Fan In,Fan Ou**t/Implemented using crossbar array, but this has limitations with regard scaling. Higher radix interconnections are being considered. (Loihi2 provides faster and higher radix interfaces). | Conductive thread, with insulation bridges to eliminate short circuits in a fabric environment , can be utilized to embed such a crossbar array design in a textile environment. Depending on the type of synapse utilized ( conductive thread resistor, SMD resistor, memristor or other) a physical connection to this synapse type will need to be considered to assess best approach in order to interconnect to the fabric-based crossbar array. |

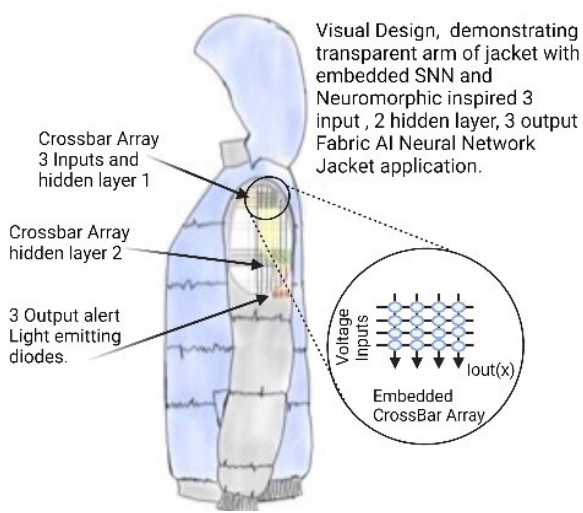

**FIGURE 8.** Conceptual Jacket design with embedded neural network using crossbar array architectural design.

From a functionality and design perspective, how can inspiration be taken and mapped to an etextiles fabric environment in order to progress towards an embedded neural network. The crossbar row/column structure design is a key element to consider, how can this architectural design be accommodated in a fabric material in order to recreate such a neural network [Fig 8]. Can embroidery based techniques and patterns using conductive thread , fabric tape or embedded woven conductive elements into a fabric environment be experimented with in order to recreate such a crossbar array type architectural structure in a fabric material. This is a key design element for further exploration and research.

### B. NEUROMORPHIC COMPUTING MONITORING/ MEASURING

*Memristors* also known as resistive switching random access memory devices that have the capability to change their resistance state and act as non volatile memories for embedded memory based devices are showing promise in the neuromorphic world as key components to implement high-density memory. Properties of memristors include small device/high density integration, low power, high speed and highly scalable( Zhang et al., 2020). New research is focusing on the potential to enable controls for resistive filament switching in synapse applications, as well as further investigation around varying memristor materials for artificial synapses with specific focus on the synaptic behaviors of organic materials, 2D materials, emerging materials (*halide perovskites*) and low-dimensional materials [33][34].The memristor is very suitable for analog based circuits as well as hardware multi-state neuromorphic applications due to its high and low resistance state. Interconnections between the neural nodes in the human brain have a joint strength represented by the synapse. Memristive synapses are ideal candidates to create an artificial synaptic device helping it mimic interconnection strengths between artificial neural nodes. A core requirement is the need to enable and alter resistance states. When we consider fabric smart material, how wearable memory can be incorporated into a fabric environment is a key element that requires extensive investigation and research. Taking inspiration from memristor-based analog memory circuits, what properties and elements need to be considered when considering the link between fabric materials and the application functionality.Analog memristors exhibit a gradual change in resistance and hence are more suitable for analog circuits and neuromorphic system applications. Bi-stable memristors act as binary memory/switches and digital logic circuits. Multi-state memistors are used as multi-bit memories, reconfigurable analog circuits, and neuromorphic circuits [32].Ames Research Center in California have implemented a method of weaving flexible computer memory into garments. This flexible memory is woven together using strands of copper and copper-oxide wires. [57] demonstrate advanced research into the development of a textile memristor using a robust fibre through an electric field assembly method that weaves the fibres into a scalable textile memristor. This exciting research era will see advances through the fusion of nanotechnology level memristor devices.

### C. NEUROMORPHIC COMPUTING COMMUNICATIONS AND ENERGY EFFICIENCY

Building on the crossbar array design, neuromorphic chip advancements look to implement energy efficient lower power consumption architectures supporting the required precision communication. In order to accomplish this research is ongoing around the design and development of smaller and multiple arrays. Such multiple arrays are emerging as either having a lateral 2D layout or a 3D vertical stacking layout. Circuit designs are required to be efficient in order to enable data flow between each layer in such 3D passive arrays. 3D memristive neural networks are taking inspiration from string stacking for 3D NAND flash (Xia et al).

From a design perspective when considering how to embed a multilayered neural network in a fabric environment, it is vital to consider key properties of the fabric as well as key design and usability functionality requirements for end users. We already touched on possible layered and woven conceptual design layout approaches in [Fig 5], but if adopting a 3D stacked layered fabric approach, how we can interconnect the layered neural node connections also need to be considered. How do we interconnect from one fabric layer to another fabric layer in an energy efficient, low power and reduced size capacity to ensure a high operational standard for the fabric based neural network. A modular fabric design-based approach with inter-changeable neural nodes and hidden neural layers may prove to be a more suitable option availing of the capability to interconnect, remove and replace neural nodes using for example snap connectors or other connector method options as described in [36].

## VI. AI TEXTILE INTELLIGENCE USE CASE EXAMPLES

Embedded AI intelligence in a fabric based environment has the potential to be applied across many sector based applications. Here we briefly provide 2 such examples in order to convey the possibilities of such an advanced fabric computing and fabric AI intelligence driven era, 1) the healthcare/rehabilitation sector application space and 2) the unmanned aircraft/drone sector where textile - driven drone control intelligence applications could be exploited.

### A. HEALTHCARE AI SMART TEXTILE USE CASE

Smart garment applications can greatly contribute towards remote monitoring, where individual and personalised healthcare provides enhanced real time assessment and early intervention. Embedded seamless AI in a textile environment adds another layer of real time intelligent wearable point of care going beyond current state of the art. The application of AI intelligence in a fabric wearable environment bring the potential for enhanced quality of life for end users. Here we provide two such conceptual end user-centric scenario examples.

- Aisling is concerned about getting a variant of COVID-19 and is looking for a new means to be able to monitor and track her general health without it impacting on her daily activities. Aisling purchased an AI monitoring package (*textile sensors and Fabric AI patch intelligence*) that can be fitted in a modular manner into her latest modular clothing garment. Aisling now has embedded artificial intelligence in her everyday clothing and can monitor her breathing and temperature, analyse the data in real time and be alerted about abnormalities that occur, allowing her to respond in an efficient manner, detect symptoms early, take a COVID test and restrict her movements if needs be.
- Jim suffers from Epileptic seizures. He was diagnosed as having tonic-clonic seizures which cause symptoms such as his muscles to stiffen, reduced breathing capability, loss of consciousness (prone to falls) and rapid jerking of the arms and legs. Due to not knowing when such a seizure can occur, Jim has become very anxious about leaving the house on this own to run errands. This has a huge impact on Jim's quality of life. Jim's doctor introduced him to the wearable modular smart AI garment with interchangeable Fabric AI embedded, that can help monitor Jim's temperature, breathing and selected arm/leg muscle Electromyography (EMG) response through embedded textile based sensors in the wearable modular garment interconnected to a Fabric AI neural network computational interchangeable patch. If Jim wears this garment when leaving the house to run his errands, the sensors will monitor his measurements, depending on the assessment completed via the Fabric AI neural network patch, this can decipher in real time if Jim is in a pre-seizure state or not. It was explained to Jim that this analysis is completed in the intelligence embedded in the core fabric and yarn of the garment and is not visible to the human eye. If a pre-seizure state is detected a simple visual alert via an embedded light alert system in the sleeve of jims garment will alert him to the pre-seizure and give jim time to try to find a safe zone in an attempt to reduce potential falls and injuries. In parallel the embedded fabric AI would also raise the alert externally to Jims carer, informing them of the pre-seizure state. This form of monitoring provided the wearer with a sense of empowerment and control.

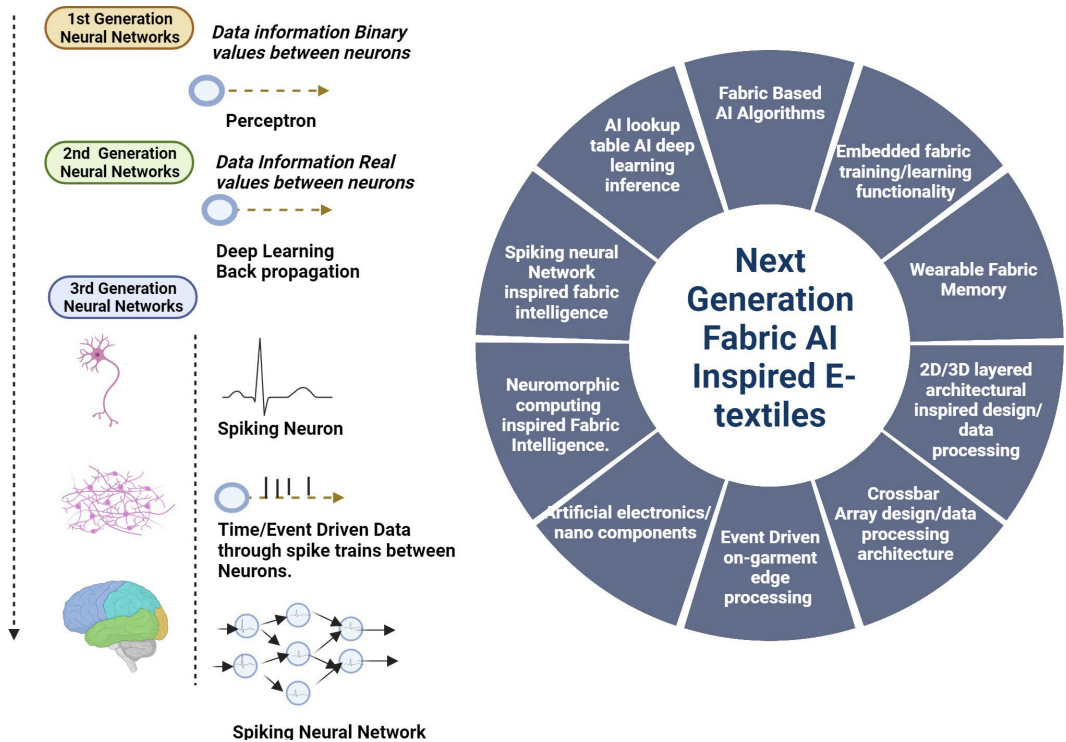

**FIGURE 9.** Core building modules and emerging technologies that are contributing towards the next smart textiles generation incorporating Fabric AI inspired e-textiles.

### B. TEXTILE-DRIVEN DRONE CONTROL USE CASE

The application of etextiles across multiple domains and sectors is gaining traction. New innovative ideas extending beyond the norm of healthcare are starting to be considered and emerging. Advances in fabric based AI intelligence open the opportunity to extend applications of etextiles fusing non traditional techniques and new technologies. An example of one such area is the application and use of a based fabric textile intelligence with unmanned Ariel vehicles (UAV). the following provides example use cases for consideration

- Taking a modular designed dynamic field programmable or Fabric AI-driven smart garment with intelligent embedded control logic functionality [55][58], opens up opportunities towards the use of such a smart garment as a control device of the UAV's based on human control activated movements linked to the smart garment triggering smart textile sensors as actuators directing the movement and control of the UAV. Such fabric AI driven haptic wearable devices can have multiple applications for varying devices providing a more seamless embedded control options for end users. This has numerous innovative applications in construction, defence and more.

### VII. CONCLUSION

Its evident that AI driven technology advancements are moving at a rapid pace. Vast research stemming from architecturally inspired specification and design properties of SNN and neuromorphic computing provide valuable inspiration towards new techniques, methodologies and designs that can be applied across to drive emerging innovations in the etextiles domain. This paper has delved into key architectural properties of SNN (*artificial neurons and synapses*) as well as neuromorphic computing (*cross bar analysis, memristors, stacked and layered design based approaches*) to stem such experimental research avenues. Table 3 and [Fig 9] provide a summarised visual of core aspects to inspire and drive new and novel innovations in the next generation fabric AI inspired etextiles.

Continued research is required in this area. Key research questions and challenges still remain unanswered hence validating the need for further research in this space. Such challenges and future research investigations include the following

- Advancements in the specification, design and verification of Fabric AI.
- Consideration around the identification and development of a Fabric AI based development language.
- Investigation into how AI algorithms can be embedded in an operational manner in a Fabric AI environment.
- In a textile environment what methods or processes can be applied to enable ML based data abstraction and processing.
- Specification and formalisation of textile driven properties to support fabric AI systems.
- The need for further investigate research around the verification of Fabric AI approaches delving into trustworthy and explainable Fabric AI.

AI technologies have developed at a much quicker pace over the past few years, its now time for the etextiles domain to embrace such advancements and build on core defined elements and properties in order to stem new and exciting research driven innovations in the etextiles domain.